\def\ie{{\it {\frenchspacing i.{\thinspace}e. }}}

\def\etal{{et al.}}
\def\simlt{\hbox{ \rlap{\raise 0.425ex\hbox{$<$}}\lower 0.65ex\hbox{$\sim$} }}
\def\simgt{\hbox{ \rlap{\raise 0.425ex\hbox{$>$}}\lower 0.65ex\hbox{$\sim$} }}

\def\msun{ \rm {M_\odot}}

\def\that{{\hat t}}

\def\zbar{{\bar z}}

\def\umin{u_{\rm min}}

\def\uas{$\mu$as}
\def\dthresh{D_{\rm th}}
\def\durthresh{t_{\rm th}}
\def\zcol{z_{\rm CoL}}

\documentstyle[aas2pp4,epsf,astrobib]{article}
\begin{document}

\title{ Astrometric microlensing as a method of discovering and characterizing
extra-solar planets}

\author{
	  Neda Safizadeh, Neal Dalal, and Kim Griest
	}
\begin{center}
{\bf Physics Department, University of California, San Diego, CA 92093}\\
\today 
\end{center}

\begin{abstract} 
We introduce a new method of searching for and characterizing extra-solar
planets.  We show that by monitoring the center-of-light motion of microlensing
alerts using the next generation of high precision astrometric instruments
the probability of detecting a planet orbiting the lens is high.
We show that adding astrometric information to the photometric microlensing
lightcurve greatly helps in determining the planetary mass and semi-major axis.
We introduce astrometric maps as a new way
for calculating astrometric motion and planet detection probabilities.
Finite source effects are important for low mass planets, but even
Earth mass planets can give detectable signals.

\end{abstract}


\section{Introduction}
Prompted by several successes 
the search for extra-solar planets
has recently intensified.  Although the radial velocity technique
(e.g. Marcy \& Butler 1998)
has yielded the best candidate planets, other promising search techniques
have been proposed or are underway:
accurate astrometry of nearby stars
\cite{sim},
direct imaging of planets,
\cite{directimage}
occultation of stars by their orbiting planets, and fine structure
on top of photometric microlensing lightcurves
\cite{maopac,gouldloeb}.

In this paper we propose a new technique that can be used to discover
and characterize extra-solar planets.  This method, using 
astrometric deviations
in the motion of the center-of-light of microlensing events, 
has several promising
features.  First, if complete astrometric coverage of microlensing alert
events were undertaken, we show that the probability of detecting any
planetary system that is present can be substantial, typically 
greater than the probability of detecting a planetary system 
when photometric monitoring alone is used.
Second, if astrometric monitoring is added to photometric microlensing
we show that in most cases the planet mass and projected orbital radius can
be determined, in many cases with just a few astrometric measurements.
Thus the degeneracies 
\citeN{gaudigould}
discussed for photometric microlensing
can usually be broken by adding astrometric information.  This planet detection
method, like the microlensing photometry method, works best for planets
at intermediate distances from their star, \ie\ in the lensing zone.
It is sensitive to planets throughout the Galaxy, and works well down to quite
low mass planets.
To be useful for this purpose, however,
the new interferometric instruments and satellites will need
to be able to respond fairly quickly to microlensing alerts, and this
need may affect their design and impact scheduling considerations.

\section{Planet Searching with Photometric Microlensing}
Gravitational microlensing is now established as a method of detecting
low-luminosity objects
\cite{nat93,eros93,ogle93}.  As a low luminosity object
passes near the line-of-sight to the monitored source, the source brightens
in a well known achromatic and time-symmetric way.  If the lens is
a faint star with an orbiting planet, a more complicated ``binary lens"
caustic structure is present and much more complicated photometric lightcurves
are possible.  For a Jupiter-like planet orbiting a Sun-like star,
the probability of a detectable deviation from the standard lightcurve
can be substantial
\cite{maopac,gouldloeb}.
Encouraged by these
calculations, several world-wide networks 
\cite{gmanfinite,planetalbrow}
have begun actively searching for such deviations using the many dozens
of bulge microlensing alerts generated yearly by the survey 
experiments
\cite{bulge45,alert,oglealert,erosalert}.

\section{Planet Searching with Astrometric Microlensing}
Microlensing of a bulge star 
by a single lens produces two images or the star separated by
several hundred microarcseconds and these images move as the 
lens passes near the observer-source line-of-sight.  
The scale of the image motion relative to the source 
is set by the angular Einstein radius
$r_e = 903 \mu{\rm as} [(m_l/\msun)\break 
(10 {\rm kpc})(1/D_l - 1/D_s)]^{1/2}$,
where $D_s$ and $D_l$ are distances to the source and lens respectively.
For example, the Einstein radius 
for a $0.3\msun$ lens at 4 kpc, with the source at 8 kpc is 550 \uas.


Recent work
\cite{gatewood,shaklan}
has shown that the Keck telescope
is capable of angular
resolution down to the sub-milliarcsecond scale.  In addition,
planned interferometers at the Keck
\cite{kecki} 
and VLT
\cite{vlti}
should have accuracies
of about 10 \uas, while NASA's Space Interferometry Mission (SIM) is planned
to have 1 \uas\ resolution
\cite{sim}.  
These instruments make resolution of astrometric
microlensing motion a feasible prospect.  
Several theoretical studies on this subject have been done recently
\cite{hoeg,miyamoto,walker,bodenshao,pacsim,hanchang,han}. 
\citeN{bodenshao}, for example, describe in detail
the predicted astrometric motion of lensed images.  They point out that the
instruments described above
will only be able to measure the motion of the center of 
light (CoL), which is smaller than the motion of the two images, but which is
nevertheless detectable and useful.  
The CoL motion
measured by an unaccelerated observer is an ellipse, whose eccentricity is a
simple function of $\umin$, the 
impact parameter of the lens relative to the source.
This is depicted in Figure 1.  
The ellipse, however, can be distorted due to the Earth'
s motion around the Sun 
\cite{bodenshao}.
Other effects such as blending, etc. 
\cite{han},
can also distort the ellipse.
Figure 1 also shows examples of
the effects of Earth motion (parallax) and blending on the astrometric ellipse.

Astrometric information is of interest because it can resolve the degeneracies
that arise in the photometric microlensing lightcurve.  
In most cases, a fit to
the photometric lightcurve gives only the event duration which is a function
of the three important physical quantities: lens mass, distance, and speed.
Addition of astrometric CoL information, particularly in conjunction with
astrometric determination of the lens parallax, can break this degeneracy
and allow determination of these physical quantities.

We note here that a planet orbiting the lens, if
positioned properly, can perturb the images and
thereby distort the astrometric ellipse.  Figure~\ref{figplanet} 
shows some examples of planetary perturbations.  
We see from Figures 1 and \ref{figplanet} 
a major difference
between planetary perturbations and the other distortions -- the planet's
effect is short in duration, while parallax and blending are important over
the entire duration of the microlensing event.  This fact enables observers
to distinguish planetary effects from the other effects.  
The planet's effects are of relatively short duration
since a planet's Einstein radius is small compared to the Einstein radius of 
its parent star.  Although the 
time duration is small, the magnitude of
the planetary perturbation can be rather large, even for Earth-mass planets,
as Figure~\ref{figearth} illustrates.
In Figures~\ref{figplanet} and ~\ref{figearth}
we do not show the parallax and blending effects.
We assume throughout that they can be independently determined.

Figure~\ref{figplanet} shows some examples of planetary astrometry 
and photometry curves
for $q=10^{-3}$ (e.g. a saturn mass planet orbiting a $0.3\msun$ star),
with $D_l=4$ kpc and $D_s=8$ kpc.
Throughout we will scale to an Einstein diameter crossing time of
$\that=40$ days, a typical value for bulge microlensing
\cite{bulge45}.
Since all our calculations are done in units of the Einstein
radius, one can easily rescale them for other durations 
or for other primary lens masses or distances.

Figures~\ref{figplanet}a  and~\ref{figplanet}b show typical non-caustic
crossing events, with deviations of about 50 \uas\ lasting several days.
Figure~\ref{figplanet}c shows the large rapid motion
associated with a caustic crossing, here about 200 \uas\ in only a few hours.
As the source crosses a caustic, a pair of images is created or destroyed
and the magnification of the these two images becomes very large as
they approach each other, causing rapid CoL motion. 

Figure~\ref{figearth} shows an example of caustic crossing for $q=10^{-5}$
and similar lens parameters as above.  This corresponds to an earth mass
planet around a $0.3\msun$ star.
As has been discussed for photometric microlensing
\cite{gaudigould,bennettrhie,wambs,griestsafi},
finite source effects are crucial when
considering small planets.
We have found the same is true for astrometric microlensing.
We have calculated finite source effects using two different methods.
Below we will discuss the method of astrometric maps, which automatically
includes this effect, and which was used to calculate the curves
in Figure~\ref{figplanet}.  
The finite source amplifications and centroid motion
depicted in Figure~\ref{figearth}b, however, were calculated using an 
ingenious technique 
devised by \citeN{gouldgau}, and expanded upon by \citeN{dominik}.
Figure~\ref{figearth}a shows the
rapid, large-scale motion of the CoL for a pointlike source.  Note,
however, that the planet's effects last a far shorter length of time than in
Figure~\ref{figplanet}c.
Figure~\ref{figearth}b shows a close-up of
the effects that finite sources can have on the astrometric planetary signal.
Depicted are the astrometric motion for stars of radius 1,3,5,9, \& 30
$R_\odot$, traversing the same path in the source plane as in 
Figure~\ref{figearth}a.
As expected, increasing the source size smears out the signal,
so that for 30 $R_\odot$ the deviation is entirely washed out.  Notice however
that for 3 and 5 $R_\odot$, the signal is easily visible, meaning that Earth
mass planets can in principle be detected using this technique.

\section{Probability of Planet Detection Using Astrometric Microlensing}

The possibility and probability of detecting planets with 
photometric microlensing
has been explored by several authors 
\cite{maopac,gouldloeb,wambs,bennettrhie,peale,sackett,sahu,griestsafi}
and was found to be substantial.
We wish to compare astrometric microlensing with these studies.
In most of these studies, 
numerous lightcurves were generated for a range of planetary masses
and projected orbital radii, 
and some simple detection criteria was established.
In a realistic experiment, fitting of the planetary lightcurve would need to be
performed and the planet mass determined, before a planet detection was
established.  Parameter extraction with astrometry will be discussed below,
but in the following section we will just use simple detection criteria 
analogous to those used in previous studies.  
We define as a detectable perturbation, any
astrometric curve that contains a deviation from the single-lens ellipse
greater than a threshold amount $\dthresh$
for a period of time at least $\durthresh$. Detection thresholds and
minimum durations needed will depend upon the interferometer or
instrument being used, so we
explore the range of detection thresholds listed in Table 1.
For comparison with photometric planetary searches, we use a photometric
detection criteria
similar to one used by \citeN{bennettrhie} and \citeN{griestsafi}:
a photometric deviation of at least 4\% for a period of at least 6 hours.
\begin{table}
\begin{tabular}{ccc} \tableline
$\dthresh$ (\uas) & $\durthresh$ (hours) & $\langle t \rangle$ (hours)\\ 
\tableline
1&	100&	370\\ 
3&	30&	160\\ 
10&	20&	50\\ 
30&	6&	15\\ 
100&	3&	13\\ 
\end{tabular}
\caption{
Detection criteria for Figure 3.  Listed are the deviation threshold $\dthresh$,
the minimum time $\durthresh$ above threshold needed for 
an event to be counted 
as a detection,
and the average time $\langle t \rangle$ above threshold of events passing the
cut for $x_p=1.3$.
\label{table1}}
\end{table}
\placetable{table1}

\section{Method of Astrometric Maps}
For a planet of mass $m_p$ orbiting a lens of mass $m_l$, the position
of the images $z=x_i+iy_i$ are found by solving the lens mapping equation 
$ z_s = z + m_l/(\zbar_l - \zbar) + m_p/(\zbar_p-\zbar)$,
where $z_l$, $z_p$, and $z_s$ are the lens, planetary, and source
positions projected to the complex lens plane
\cite{witt90}.
For a given $z_l$, $z_p$, and $z_s$ this equation is a 
5th degree polynomial in $z$
and has 3 or 5 physical solutions corresponding to 3 or 5 images.  The
magnification $A_i$ of each image is the reciprocal of the Jacobian determinant
of this mapping evaluated
at the image position $z_i$, and the total magnification is 
just $A = \sum |A_i|$.
The center-of-light (CoL) can be found from $\zcol = \sum z_i |A_i|/A$,
and it is this quantity that is measured by an interferometer.
To calculate the CoL motion, one computes $\zcol$ along a given source
trajectory through the lens plane.  This method was used in producing
Figures \ref{figearth} and \ref{figfit}.

To generate a large number of CoL curves and to get an overview of
the astrometric motion for all possible trajectories, one can alternatively
produce an ``astrometric map" of the source plane using the
ray tracing technique
(Schneider, Ehlers, \& Falco 1992, p. 303).
This is analogous
to the magnification maps 
\cite{wambs,griestsafi}
previously used in calculating photometric microlensing probabilities.
Using the above lens equation, one simply ``shoots" many photons from all 
pixels of the image back to the source plane, and records the photon
weighted $x$ and $y$ deflection.  The resulting $x$-shift
and $y$-shift maps therefore show the CoL shifts for a source at each
position in the source planet.  
A CoL curve is created by simply tracing a trajectory
through the 2 maps.  To calculate the probability of detecting a planet
via astrometry, we consider a complete set of 
trajectories (with $A_{\rm max} \geq 1.6$) and find the fraction of these that
pass the thresholds listed in Table 1.
We then repeat this procedure for various values of $x_p$, the projected
planet-lens separation, and $q$, the planet-lens mass ratio.
We note that since binning of the photons
is necessary, the finite source-size effect is automatically taken into 
account.  Different source sizes are easily handled by convolving the
maps with a round star-sized, limb-darkened kernel.

The results for a saturn ($q=.001$) mass planet around 
a $0.3\msun$ star, with the system at a distance of 4 kpc, the
source star at 8 kpc are shown in Figure~\ref{figprob}.  The
astrometric maps used to calculate this figure 
are themselves interesting, and these will be
presented in a subsequent paper
\cite{safigriest}.
We see that the probabilities are substantial for a wide range of
planet-lens separations.  
For the case shown, $x_p=1$ corresponds to 2.2 A.U., so the lensing zone
($0.6 \leq x_p \leq 1.5$) corresponds to 1.3 - 3.3 A.U.

From Figure~\ref{figprob} we see that
probabilities range from over 70\% over the entire
lensing zone for the optimistic SIM
threshold of 1 \uas\  deviation down to $\sim 50\%$ for a 3 \uas\ threshold.
For the 10 \uas\ accuracy expected from the Keck and VLT 
interferometers, the probabilities are 20\%-40\%,
while they are below 20\% for the 100 \uas\  
accuracy reachable with current technology.
For comparison, 
the dashed line at around 30\% probability shows detection probabilities
using photometric lightcurves, and a 4\% deviation criteria.
For a jupiter mass planet the probabilities are $\sim 15$\% higher,
while for a 10 earth-mass planet they are smaller by a large factor.
If the new generation of astrometric instruments do as well as we have assumed,
they should be excellent planet detectors.

Of course these probabilities should be averaged over planetary
orbit orientations, and a complete exploration of the planetary mass and 
semi-major axis parameter space is necessary.
We will present these results elsewhere, as well as a study of the
the effect of the finite
source size on small (e.g. Earth mass) planets and the loss of signal when 
the astrometric deviation is found by subtracting a fitted astrometric 
curve rather than
using a priori knowledge of the single lens curve.
Convolving the saturn mass maps with a giant star kernel results 
in almost unchanged probabilities, so finite
source effects are not important in the saturn mass case presented here.

\section{Extraction of Planetary System Parameters}

Microlensing has both disadvantages and advantages with respect to
other methods of planet detection.  The main disadvantage is that the
entire signal is a short duration deviation on a lightcurve, so
further exploration, or even attaining additional information concerning
the detected planet, is probably impossible.  The main advantage of 
microlensing is that large numbers of planets can be found throughout
the Galaxy and therefore statistics can be found on the frequency
of planetary systems and the distribution of planetary masses and
semi-major axes.  Most other planetary search methods
are restricted to a small sample of nearby stars. 

Since no follow-up information can be obtained,
in microlensing it is clearly important to get as much information
as possible during the brief planetary deviation from single-source
lensing.  Thus, adding the two additional astrometric curves
($\Delta x$ and $\Delta y$) should be extremely valuable.
In addition, 
a convincing demonstration of a microlensing planet detection
will require determination of the planet mass. 
This will be done via fitting the photometric and/or
astrometric curves to extract the planetary system parameters.

To test this hypothesis,
we have constructed a simple algorithm for fitting data sets.
We generated numerous simulated noisy data sets, and attempted to
fit them using our automated routine.  There are 8 physical parameters,
5 of which describe the primary lens and trajectory, 
and three of which describe the planet.  The 5 primary lens parameters are
the primary lens Einstein radius, the minimum impact parameter $\umin$, 
the Einstein diameter crossing time ${\hat t}$, the trajectory
direction, and the 
time of closest approach.  The three planetary parameters are the planet-lens
mass ratio $q$, the planet-lens projected separation $x_p$, and the planet's
angular position relative to the lens.

Since the planet's effects on the astrometric
motion and magnification are usually perturbative and short-lasting,  
we first fit the data sets to a single isolated lens, to extract the primary
lens parameters. We then hold those parameters fixed while we fit for the
planetary parameters.  Given this initial parameter set, we then
do a fit in the full 8-dimensional parameter space.  
We find that our automated
algorithm can almost always find a good estimate of the 
planetary parameters, for planetary deviations that are significant compared
to the added noise.
In these tests we have ignored planet-lens orbital motion and Earth
parallax.  Orbital motion adds several more parameters and will usually
be small during the short planetary deviation.
We note, that if in addition, astrometric parallax
\cite{bodenshao}
is found by monitoring the event long before or after the planetary deviation,
then the physical size of the lens Einstein radius can be found
and the physical planetary mass and separation can be determined.

Figure~\ref{figfit}  shows an example data set and overlaid
best fit.  Note that there are only 8 measurements taken during the
planetary perturbation, and yet full parameter extraction was possible.
It is important to note that the ability to recover physical
parameters depends greatly on the noise in the signal.  
The example of Figure~\ref{figfit} uses $q=10^{-3}$, and
$x_p=1.3$, with Gaussian noise of
$\sigma=5$ \uas\ added to each astrometric data point and 5\%
Gaussian noise added to each photometric data point.  If we define
a signal/noise by the size of the planetary deviation divided by
the astrometric noise, then we find we are able to recover
the input parameters reliably for a S/N greater than 4 or 5
with 6 or 7 data points during the planetary deviation.
For lower S/N,
we find evidence of the 
degeneracies discussed by \citeN{gaudigould} starting to appear,
and for even smaller S/N, parameter extraction at all becomes problematic.
A more complete exploration of the effect of noise on
the ability to recover parameters will be presented elsewhere 
\cite{dalalgriest}.

Besides noise, the sampling rate can also limit parameter extraction.
Since time on the new interferometers will be very valuable and hard to obtain,
sampling at first may be poor, so
it is important to know what happens
when the assumption of complete astrometric coverage is relaxed.
One would like to know the minimum number of astrometric
measurements and the minimum accuracy that is needed to extract
reasonable values of the planetary parameters.  This work is in
progress.

As an example of why astrometric data is so helpful, consider the
``major image/minor image" degeneracy discussed by \citeN{gaudigould}.
For a single lens there are two images.
The minor image is dimmer and inside the critical curve (Einstein
radius), while the major image is always brighter and outside.
It is not easy to tell from photometry alone whether the planet is perturbing
the major image or the minor image.
During a planetary perturbation the bulk of the astrometric deviation
is due to the changing image brightness rather than the image motion,
so there should be a correlation between the photometric and astrometric
deviations.
For example, if the major image is perturbed and brightens, then the CoL motion
will be forced outside the single-source ellipse, and the photometric
deviation will be positive (total magnification increases).  However,
if the minor image is perturbed and brightens, the CoL motion will be
forced inside the ellipse with a positive photometric deviation.
Opposite correlations exist for perturbations that cause an image
to dim, so by simply
correlating the CoL motion with the photometric deviation one easily
can tell which image is being perturbed by the planet.  
See Figure~\ref{figplanet} for examples.
This correlation, 
greatly helps in determining the position of the planet, and is used in
our algorithm.
The above correlation works when the perturbation is due to an image nearing
the ``planetary caustics", but we find we can also distinguish the positions
for close encounters of the ``central caustic"  (see \citeN{griestsafi}
for discussion of this terminology).

The most crippling degeneracy discussed by \citeN{gaudigould} 
arose due to finite
source effects.  We have begun exploring finite source effects,
and will present our detailed results elsewhere (Dalal \& Griest 1998).
For the saturn mass planets of Figure~\ref{figprob}, 
these effects are small and our
fitting algorithm seems to work well.

\section{Summary and Discussion}

By adding astrometric measurements to ongoing photometric monitoring
of microlensing alerts towards the Galactic bulge, one can in many
cases, determine the planetary mass and projected planet-star separation.
We feel this is an important reason that SIM pointings should be made
towards microlensing alerts.  SIM is still in its design phase, and it
is therefore important that the satellite software and hardware be capable of
responding quickly to ``planetary alerts" from the photometric microlensing
planet search teams.  We note (from Table 1) that the average time
above 1 \uas\ deviation for a saturn mass planet is about two weeks.
On average the time spent above 10 \uas\ deviation is about two days,
while a 30 \uas\ deviation lasts 10 hours.
For smaller mass planets the average durations are smaller and require
faster response \cite{safigriest}.

We have also shown that, using our criteria, we can expect to detect
and characterize a substantial fraction of planets that happen to 
orbit lenses,
if reasonably complete astrometric monitoring is undertaken.
The microlensing surveys typically produce more than 100 events a year, so
microlensing has the potential to detect and characterize
large numbers of planets.
Using astrometric and photometric microlensing, 
we could for the first time begin to
measure the distribution of planet masses and orbital distances throughout
the Galaxy.

\acknowledgements

We thank Andreas Quirrenbach for many helpful discussions.
This work was supported in part by the Department of Energy under grant
DEFG0390ER 40546, and by a Cottrell Scholar award from Research Corporation.


\begin{figure}
\plotone{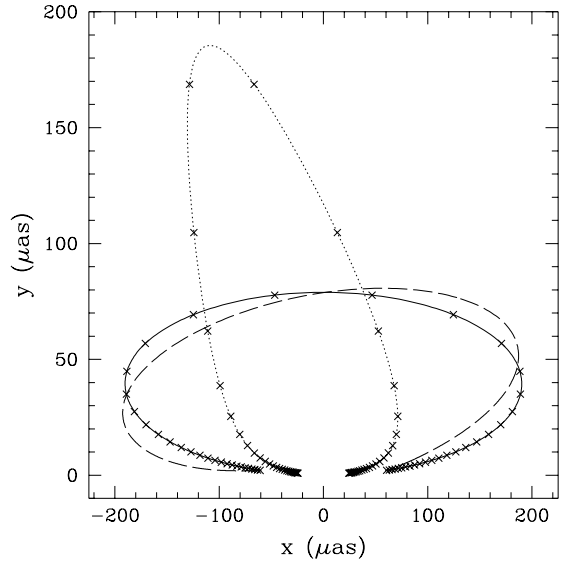}
\caption{
Non-planetary astrometric curves.  
The solid line shows a simple
single-lens curve with $\umin=0.3$, and $\that=40$ days.
The curve is plotted over one year, with x's marking each week,
so only the 5 or so weeks at largest $y$ have magnification greater than 1.34.
The dashed line shows the same with an example parallax 
effect included, while the dotted line shows the effect of blending
(blend fraction $f_b=60$\%).
\label{figsingle}}
\end{figure}

\begin{figure}
\plotfiddle{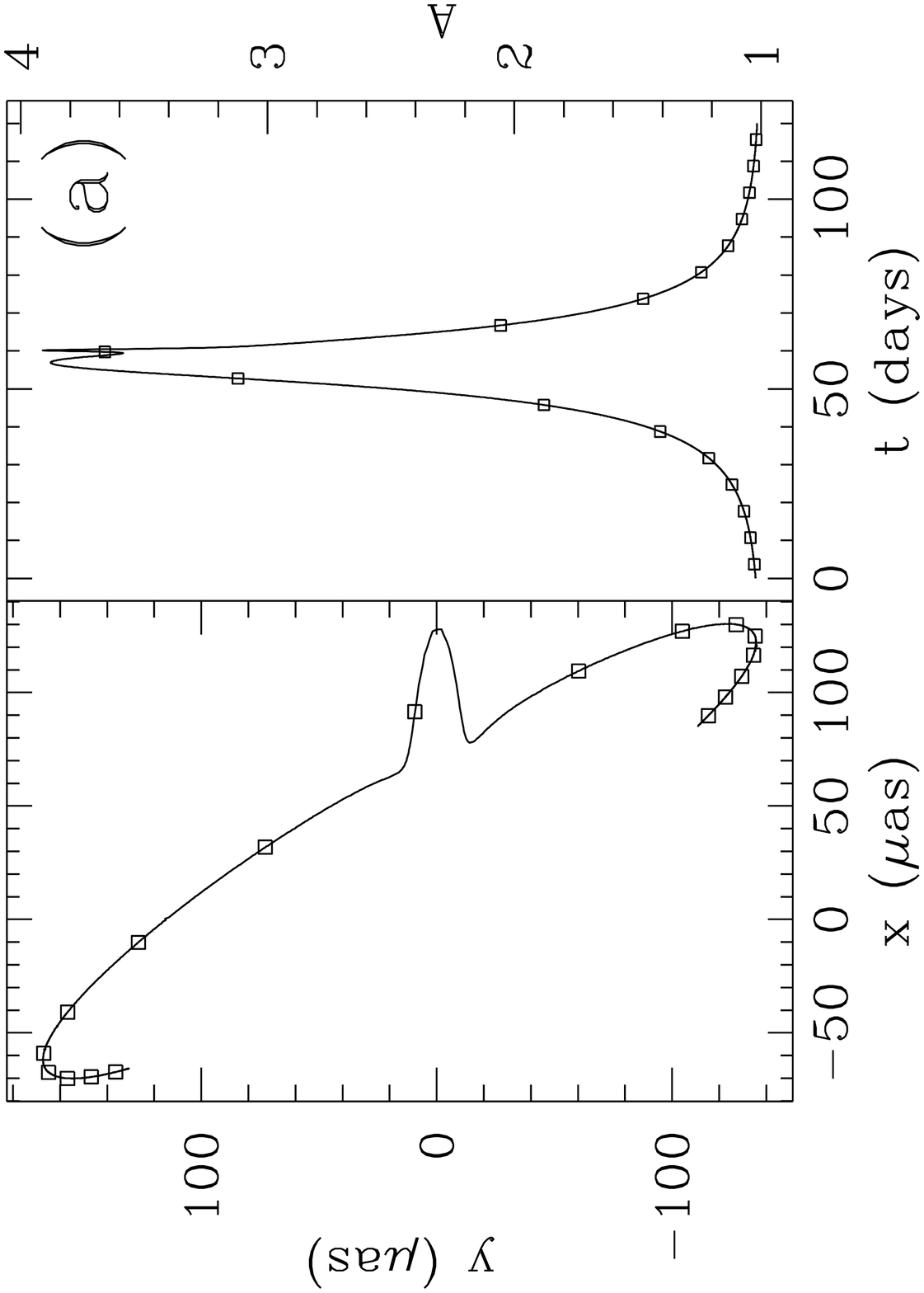}{5.5cm}{-90}{30}{30}{-122}{200}
\end{figure}
\begin{figure}
\plotfiddle{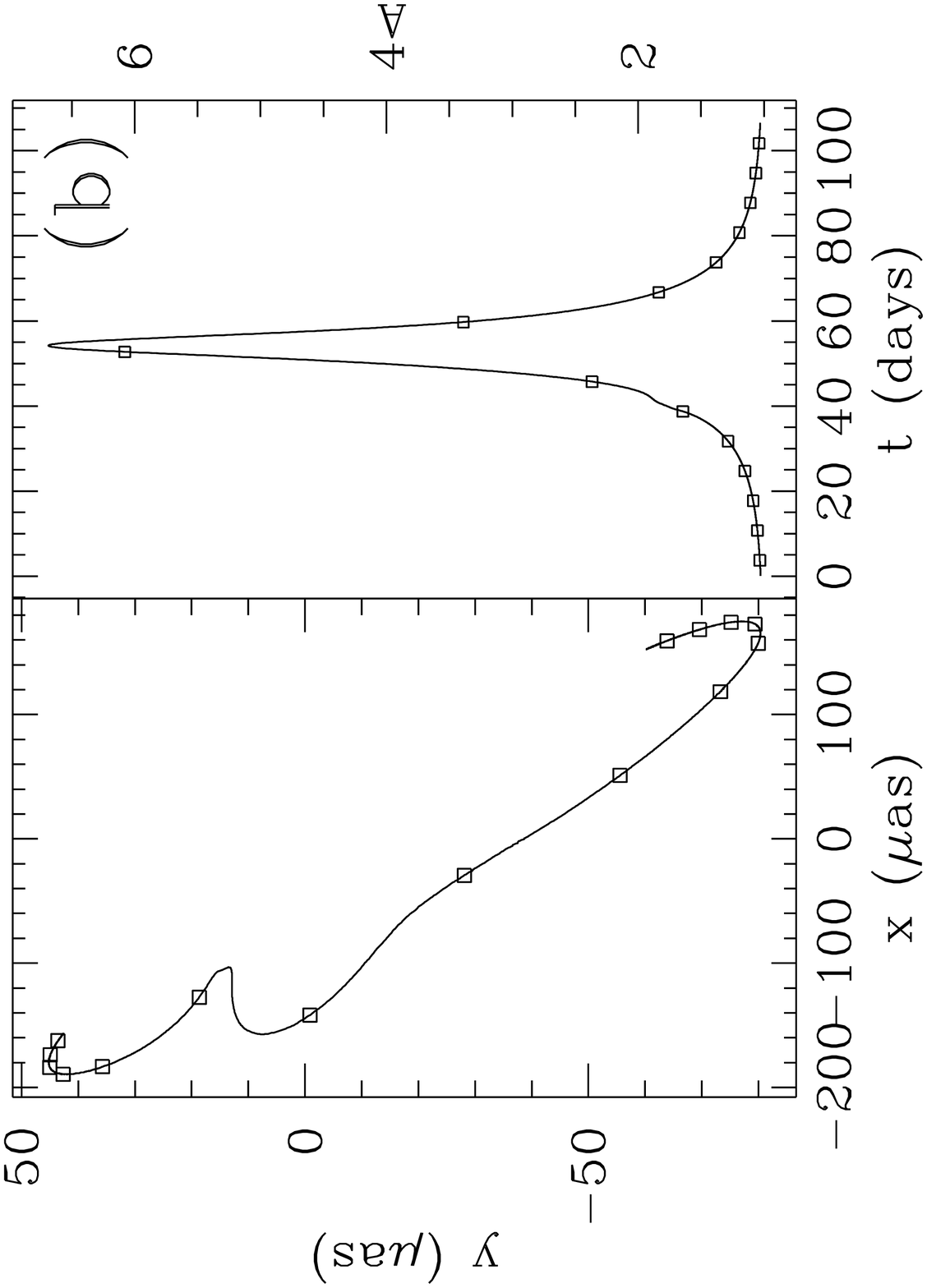}{5.5cm}{-90}{30}{30}{-122}{200}
\end{figure}
\begin{figure}
\plotfiddle{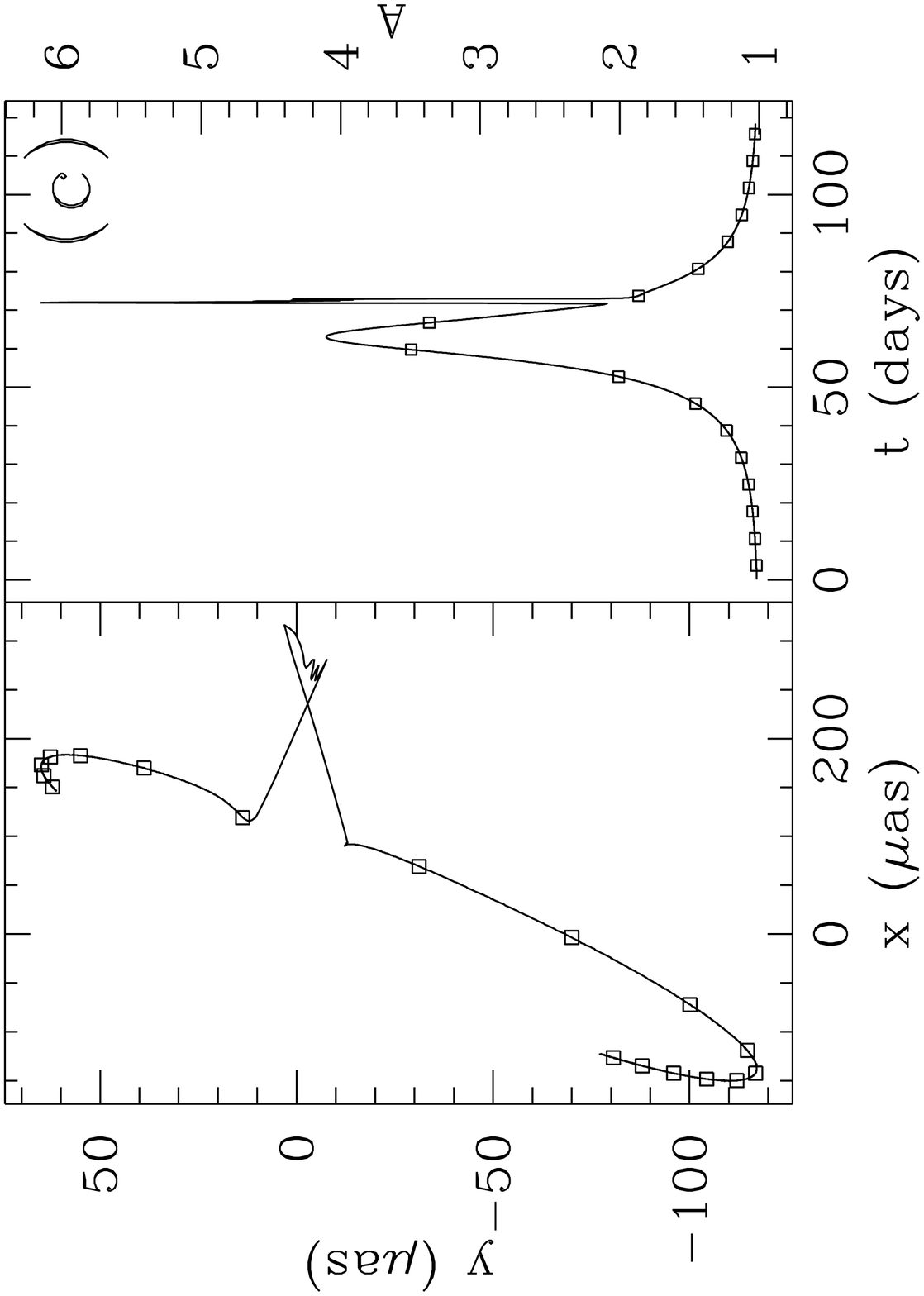}{4.5cm}{-90}{30}{30}{-122}{178}
\end{figure}
\begin{figure}
\caption{
Some examples of planetary astrometric and photometric curves.
All examples assume $q=10^{-3}$, with a primary lens Einstein radius
of 550 \uas, corresponding to a saturn mass planet.
Panel (a) has $x_p=1.3$, panel (b) has $x_p=0.7$, while panel (c) shows
a caustic crossing event with $x_p=1.3$.
The time axis is scaled so that $\that=40$ days, and squares are
plotted one per week, so the durations of the deviations are of order one week.
\label{figplanet}}
\end{figure}
\break
\vfill
\begin{figure}
\plotfiddle{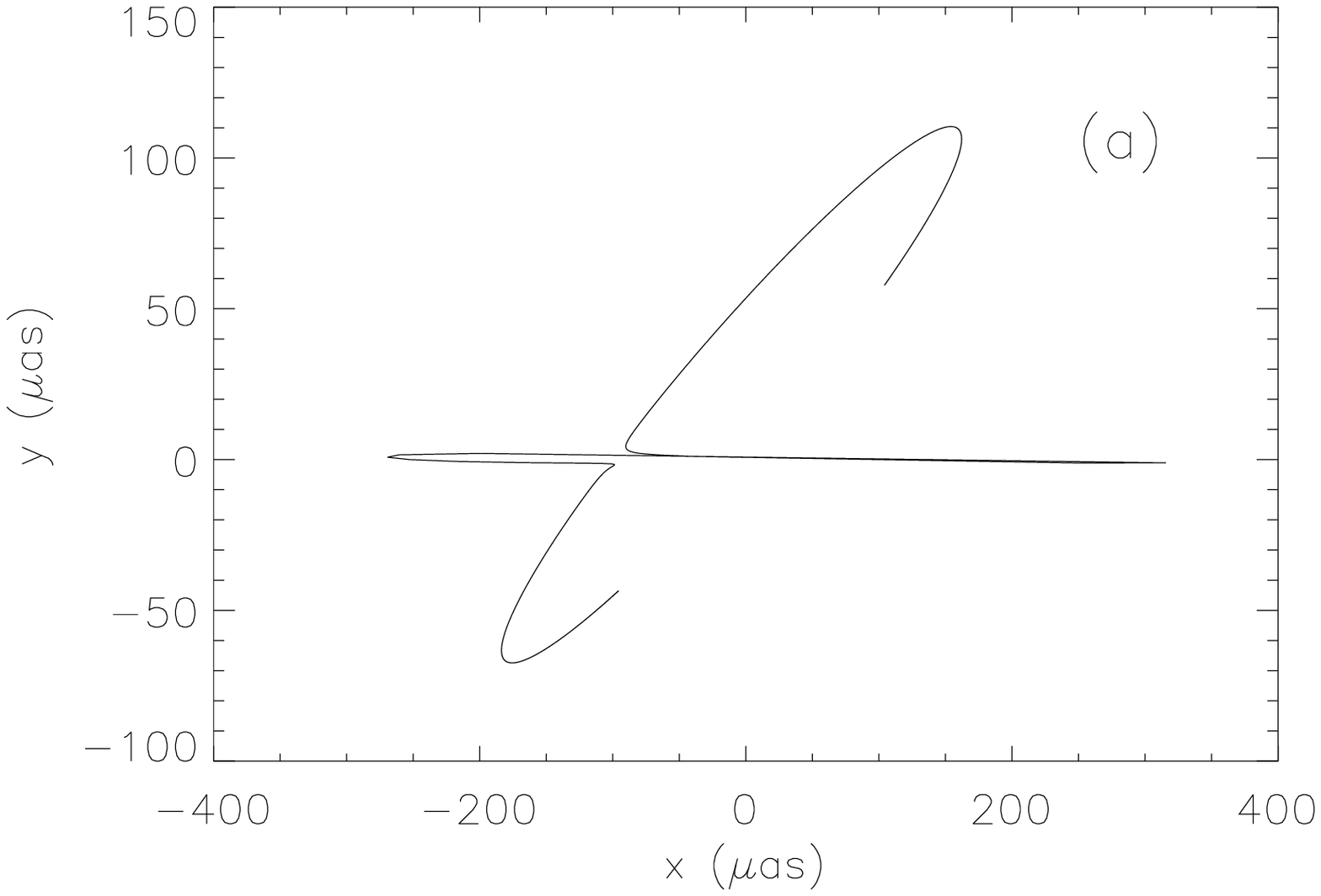}{4.7cm}{0}{50}{50}{-132}{-10}
\end{figure}
\begin{figure}
\plotfiddle{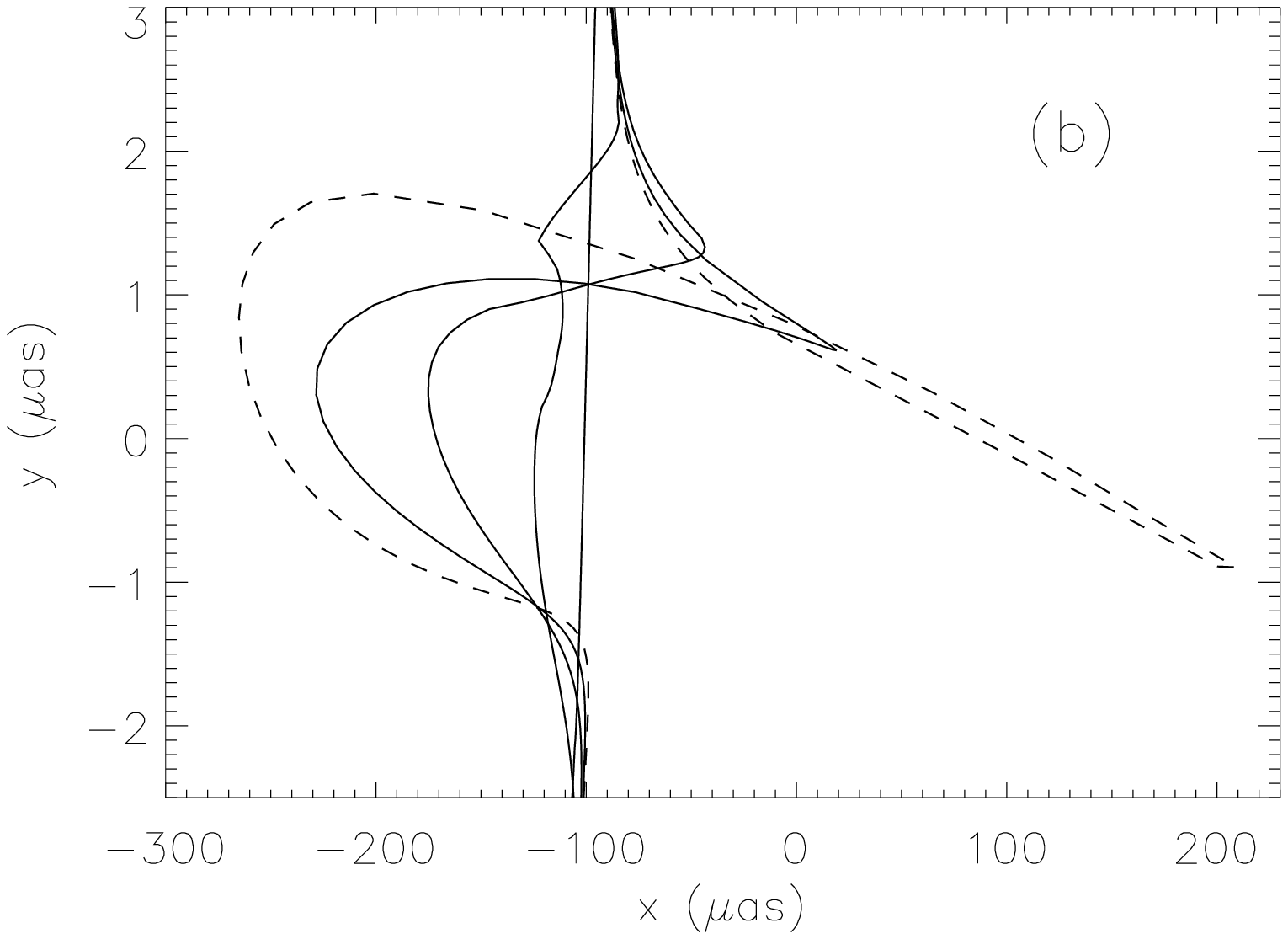}{4.7cm}{0}{50}{50}{-132}{-10}
\caption{
Astrometric motion for Earth-mass caustic crossing.
Panel (a) shows the center-of-light motion 
for a point source, crossing a caustic
associated with an Earth-mass planet at $x_p=0.825$.  The primary
lens is $0.3 M_\odot$ at $D_l=4$ kpc, and source at $D_s=8$ kpc.
Panel (b) shows a
close-up view of the planetary deviation, with finite-size source.
The dotted line plots the CoL motion for a 1 $R_\odot$ size source.  The
solid lines depict the CoL motion for more realistic sizes typical of
Galactic bulge stars, respectively 3, 5, 9, \& 30 $R_\odot$.  Note the
extreme anisotropy of the axes on the graph.
For $\that=40$ days the duration of the deviation is about 20 hours, with
the center of the source spending roughly 90 minutes inside the caustic.
\label{figearth}}
\end{figure}

\begin{figure}
\plotone{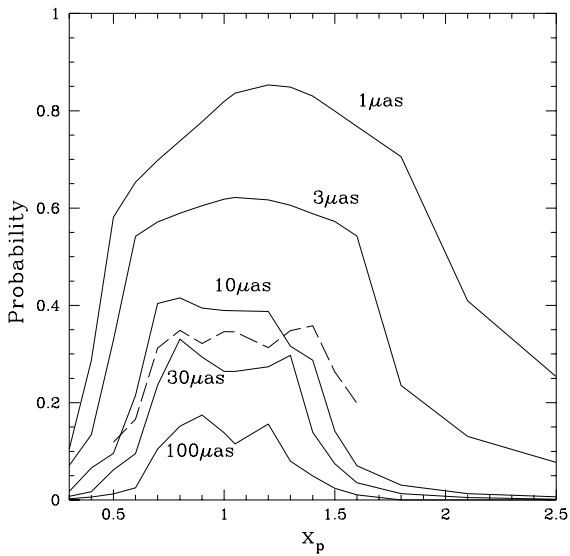}
\caption{
Probability of planet detection via astrometry versus $x_p$, the projected
planet-lens separation in units of the lens Einstein radius.  
A planet-lens mass ratio $q=10^{-3}$ 
and duration (Einstein diameter crossing time)
$\that=40$ days were assumed.  
For a source distance of 8 kpc, lens distance of 4 kpc, this corresponds to
a saturn mass planet around a $0.3\msun$ star, or a jupiter mass planet
around a $1 \msun$ star.
The solid lines are labeled
by the thresholds used (detailed in Table 1).  The dashed line shows
the probability for photometric detection with a threshold of 4\% deviation
over 6 hours. 
The 1 \uas\  and 3 \uas\ thresholds are only relevant for 
the SIM (Unwin, \etal\ 1997),
with 1 \uas\ being the SIM design goal. 
The 10 \uas, and 30 \uas\ thresholds
would be easy with SIM and perhaps possible with the Keck or VLT 
interferometers (Colavita, \etal\ 1998;  Mariotti, \etal\ 1998).
The 100 \uas\  threshold may be possible with existing
instrumentation on the Keck (Gatewood 1998;  Provdo \& Shaklan 1996).
\label{figprob}}
\end{figure}

\begin{figure}
\plotfiddle{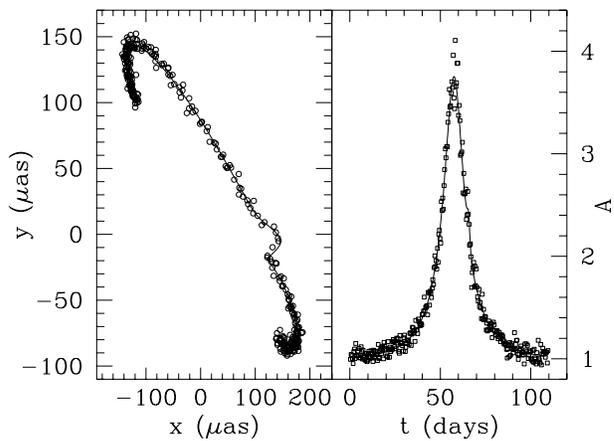}{5cm}{-90}{30}{30}{-122}{178}
\caption{
Recovery of planetary parameters using a joint photometric
and astrometric fit.  The dots show simulated data for $q=10^{-3}$,
$x_p=1.4$, $\umin=.27$, with
5 \uas\ noise added to the astrometric data and 5\% noise added to the
photometric data.  The solid line is the best fit line found by our
automated algorithm, with recovered parameters basically equal to those above.
Scaling the time axis so $\that=40$ days, the time between each data point
is about 9 hours.
\label{figfit}}
\end{figure}

\end{document}